# High-pressure Ion Trap

Evgeny V. Krylov[1]

## Abstract

A High-Pressure Ion Trap operating at pressure ~1 Torr is a core component of the portable hand-held mass-spectrometric gas analyzer. A comprehensive mathematical model of the HPIT is described in this paper. The influence of the instrumental parameters (gas composition, pressure, and temperature; applied voltages; ion trap size and geometry) and ion properties (mass, diffusion, and mobility) on the ion trap analytical parameters (mass-spectral peak position, height, and width) is examined. The model explained the difference between a high-pressure and regular low-pressure ion trap.

## Introduction

The Quadrupole Ion Trap is an extraordinary device that functions both as an ion store in which gaseous ions can be confined for some time and as a mass spectrometer of considerable mass range and variable mass resolution. While such an accurate and sensitive instrument exists in bulky form in many analytical laboratories, there is a high demand for a smaller, handheld mass spectrometer for remote and widespread deployment. This demand has stimulated researchers to seek ways to miniaturize and simplify an MS system. The main obstacle on that path is a vacuum system, which is normally large and heavy and consumes much power to provide the required pumping speed and pressure. However, if it is possible to operate an MS at a higher pressure, the pumping requirements are drastically reduced. For example, if the operating pressure is more than 1 Torr, then a turbomolecular pump can be replaced by a robust and reliable scroll pump or even a diaphragm pump at 10 Torr.

The scaling rules predict that shrinking MS dimensions allows operation at elevated pressure and keeps the same mass range and resolution. The quadrupole ion trap is particularly conducive to scaling. The scaling rules for the ion trap in the application to the operating pressure, voltage, and frequency were examined in Ref[1]. They concluded that for the ion trap of characteristic size 20 µm, the maximum operating pressure should be greater than 1 Torr. However, the same team reported[2] a High-Pressure Cylindrical Ion Trap that operated at pressures up to 1.2 Torr and had dimensions of $r_0$ = 500 µm and $z_0$ = 650 µm, which was commercialized by the Boston-based company 908 Devices[3] later on under trademark MX-908. Such a huge discrepancy between predicted and real parameters evidences that standard on trap theory is not directly applicable to High-Pressure Ion Trap (HPIT) and requires a new approach, which is presented here in the **HPIT theory** section.

Before moving forward, let's make some general notes. Ion Trap (IT) miniaturization applies some technological limitations to the mini-IT design compared to the regular large-scale IT:

- relative mechanical accuracy is much lower;
- relative ejection holes size is much larger;
- only cylindrical or rectilinear geometry may be used;
- effective RF voltage is smaller than the applied one because the electric field is non-quadrupolar;
- the parallel operation of the mini-ITs (mini-IT array) should increase device sensitivity but decrease resolution (because the mini-ITs are not identical).

## Ion Trap theory

Let's begin with the low-pressure ion trap theory. IT as a mass analyzer implements mass selective instability by raising the radio frequency trapping voltage (V) so that ions are ejected in increasing mass order. The standard theoretical approach[4] is based on the solution of Mathieu's equation of ion motion in a quadrupolar AC-RF electric field. The solution is represented as a Stability Diagram (SD), which maps the region of the stability of the ion trajectory inside the Ion Trap in dependence on the experimental conditions: trap size and geometry (e.g. $r_0$, $z_0$. for the Quadrupolar Ion Trap), ion mass (m), amplitude and frequency of the AC-RF voltage (V, f), and DC bias voltage (U) applied to the IT electrodes. The position of the operational point on the SD determines whether an ion is trapped within the IT or ejected. SD is defined by the function β (stability condition is 0<β<1) plotted in the dimensionless coordinates (q; a), which are used for the normalization of the IT dimensions, ion mass, and RF frequency. For the Quadrupolar Ion Trap (QIT), equations for the dimensionless coordinates are $q_z = \frac{8eV}{m(r_0^2+2z_0^2)\omega^2}$, $a_z = \frac{-16eU}{m(r_0^2+2z_0^2)\omega^2}$, where e is the elementary charge; V is the zero-to-peak amplitude of the AC-RF voltage, and U is the DC voltage applied to the electrodes; ω=2πf is the angular RF frequency, $z_0$ is the half distance between the cap electrodes, $r_0$ is ring electrode radius; and m is the ion mass. Stability Diagrams for the regular low-pressure ITs are more or less consistent: experiment, theory, and simulation agree.

---

[1] Author e-mail: great418@gmail.com

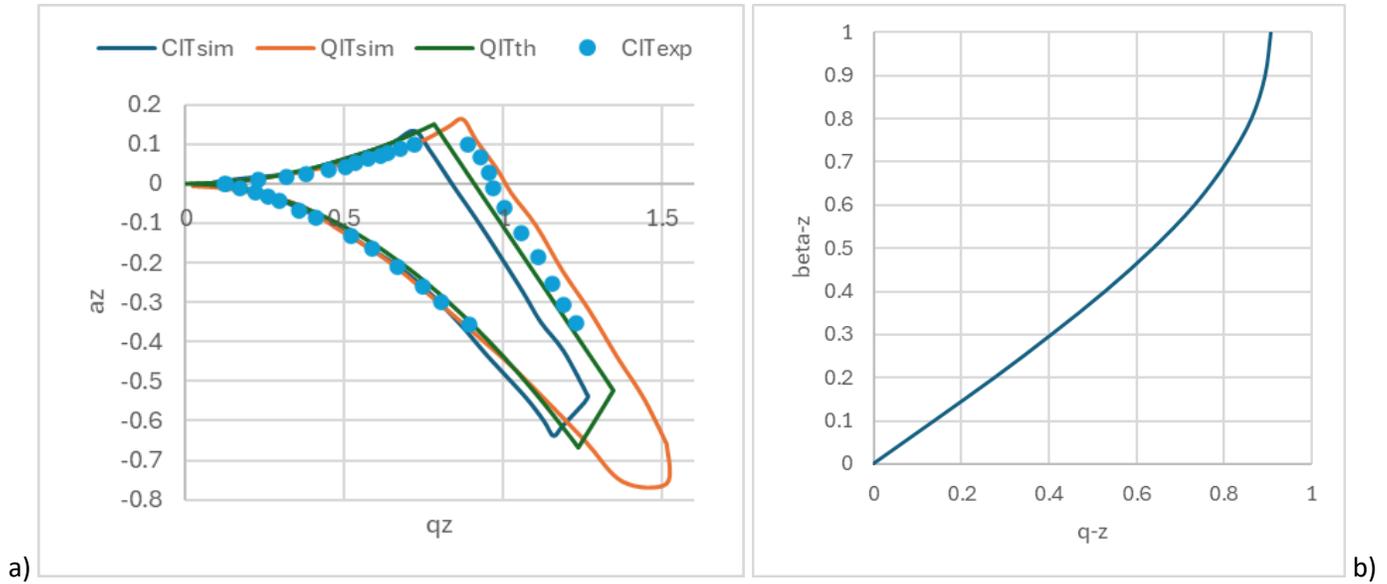

a) b)

Figure 1. a) Low-Pressure Ion Traps Stability Diagrams: QITth – Quadrupolar Ion Trap theory; CITexp – experimental data for the Cylindrical Ion Trap; CITsim and QITsim - simulations; b) Function $\beta_z(q_z)$.

Trapped ion movement could be approximated as oscillating in a parabolic pseudopotential well (PPW). Even though the IT's electric field oscillates rapidly with time, its time-averaged effect is essentially a pseudopotential with a restoring force pushing ions toward the trap's center (IT confinement or trapping phenomenon). The PPW approximation simplifies and quantifies the complex motion of ions trapped in an oscillating electric field. For low q pseudopotential trapping potential is $U_C = \frac{Vq_z}{8}\frac{z^2}{z_0^2}$.

For the full range of q PPW potential is $U_C(q,z) = F(q)\frac{m(r_0^2+2z_0^2)\omega^2}{64e}\frac{z^2}{z_0^2} \sim Fz^2$, where F(q) is the **normalized Pseudo-Potential Well depth** for the 0<q<0.908. PPW depth across the stability diagram depends on the stability parameter $\beta_u$, which is defined precisely by a continued-fraction expression without an accurate analytical solution. There are two well-known approximations of the PPW depth for low and high q-values: $\frac{\beta_u^2 V}{2q_u}$ for q<0.4 and $\frac{(1-\beta_u^2)V}{2q_u}$ for q>0.8. Also, there is a matrix method of solving the Hill and Mathieu equations[5], which can compute SD boundaries and PPW depth for the full range of q. This method relies on creating a series of 2×2 matrices describing ion behavior during a series of small potential steps that comprise one period of the driving waveform. These matrices may be multiplied sequentially over the whole RF period to define whether ion motion is periodic and stable (the stable region is bounded by curves where the stability parameter, β, has a value of 0 or 1). This method allows to calculate the PPW depth[6] at any point within the stability region: $\frac{V}{\pi^2 q_u}\left[arccos\left(\frac{|tr(M_u)|}{2}\right)\right]^2$, where $tr(M_u)$ is trace of the multiplied matrices.

In the case of the HPIT, it seems to be more realistic to approximate the normalized PPW depth pseudopotential in the full range of $q$ as $F(q) = q^2\left(1 - \left(\frac{q}{0.908}\right)^Q\right)$, where $Q$ is a constant coefficient.

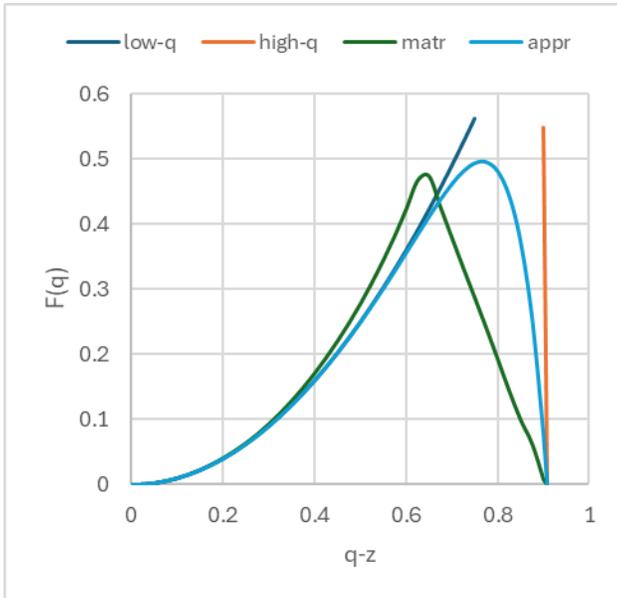

Figure 2. Normalized PPW depth, F(q), for the full range of *q* (*low-q* – approximation for q<0.4; *high-q* – approximation for q>0.8; *matr* - matrix method calculation; *appr* – proposed approximation for Q=11).

An Ion Trap can confine a limited number of ions. **Ion Trap Capacity** is the maximum number of ions that can be confined in the ion trap and still achieve a set level of performance. There are three different Ion Trap Capacities:

- The storage limit is the maximum number of ions confined within the ion trap because Coulombic forces push any further ions out.
- The isolation limit is the number of ions trapped, allowing a given isolation efficiency. The isolation limit is determined by ion/ion coupling, which prevents resonant ion selection.
- The spectral limit is the number of ions trapped while achieving the desired resolution, scan speed, and mass accuracy.

Notice that the isolation limit is typically roughly 10-fold lower than the storage limit, and the spectral limit can be 50 times smaller than the isolation limit.

The spectral limit is the most important for the mass spectrometric operation, but the storage limit is the easiest to estimate. It is reached when the repulsive electrostatic potential within the ion cloud balances the trapping pseudopotential. Poisson's relation yields maximum density of singly-charged particles that can occupy an ideal QIT[7], $\frac{3m\omega^2}{64\pi e^2}q_z^2$. For the regular QIT (pressure P=10$^{-6}$ –10$^{-3}$ Torr, ring electrode radius $r_0$=1cm, trapped ion mass m=40 Da, operating q=0.38, RF frequency f=0.762MHz,) ion density is 0.45x10$^7$ cm$^{-3}$, which is close to the experimental value 1.04x10$^7$ cm$^{-3}$ reported in Ref.[8]. The spectral limit for the regular QIT is ~2x10$^4$ cm$^{-3}$. An experimental laser tomography study shows[9] that the ion density distribution inside the trap is quasi-Gaussian with a characteristic radius s~ 0.06 cm for the ion trap of $r_0$=1cm radius. So, the corresponding numbers of the trapped ions are ~10$^4$ for the storage limit and ~20 for the spectral limit. Assuming that relative ion cloud size (s/$r_0$~0.06) stays the same for any trap radii, the number of the ions confined inside the miniature IT ($r_0$=500μm) can be estimated as ~2 for the spectral limit. An electron multiplier is necessary to detect ejected ions. But any electron multiplier requires a high vacuum to operate. It seems to make meaningless the whole idea of the miniature High-Pressure Ion Trap.

Standard IT theory was developed for the ion moving in a vacuum under the action of the electric field and didn't take into consideration the strong ion-neutral collisional scattering present in the HPIT. It doesn't answer whether the ion trajectory is stable against disturbances, which is acceptable if the collisional scattering is negligible (i.e., in a vacuum). If the collisional scattering is important (as for the HPIT), the standard theory (solution of Mathieu's equation) doesn't adequately describe the ion movement in the IT.

Authors of the Ref.[10] add a drag term $\zeta \sim PM/m$ (M is neutral gas mass, P is the gas pressure) in the differential equations of motion of a trapped ion $\ddot{u} + 2\zeta\Omega\dot{u} + \Omega^2 u = \frac{eE}{m} Cos(\omega t)$ (u is ion position; e is the ion charge; E is RF field amplitude; $\Omega$ is ion oscillation eigenfrequency) and by the substitution of variables, convert this equation into the standard Mathieu equation of motion. Drag term results in the shift of the SD boundaries inward by factor $\sim \zeta^2/\omega^2$. The computational method reported in Ref.[11] predicts that pressure dependence of the ejection point has a minimum between 0 and 5 Torr and then increases way beyond 0.908 (up to $q_z$=1.5). These approaches mix two phenomena: ion movement due to the electric field and collisional ion scattering, which seems methodically incorrect.

## HPIT theory

HPIT operates at a pressure of about 1Torr, so the mean time between two consecutive collisions of the ion and neutral is estimated as $\frac{\mu K}{e} \approx 1\mu s$, where K is the ion mobility coefficient, $\mu = mM/(m + M)$ is reduced mass (m and M are ions and neutral mass); e is the elementary charge. There are two characteristic times for HPIT: RF period ~0.1μs and trapping time ~1ms. Phenomena related to the RF field (confinement, ejection) should be considered as a vacuum motion (described in the **Ion Trap theory** section), ion energy gain in the electric field and ion losses due to the scattering should be considered a through-gas movement. Thus, the problem of the ion trapping at high pressure can be reduced to the problem of the ion moving through a gas in the potential well. Ion movement through a gas is characterized by a mobility coefficient K, so the ion drift velocity is $v_d$=EK, where E is the electric field strength. Ion scattering is characterized by the diffusion coefficient D, so the ion cloud spreads in space according to the phenomenological Fick's first law: $j = -D\nabla n$, where j is ion flow; n is ion cloud density.

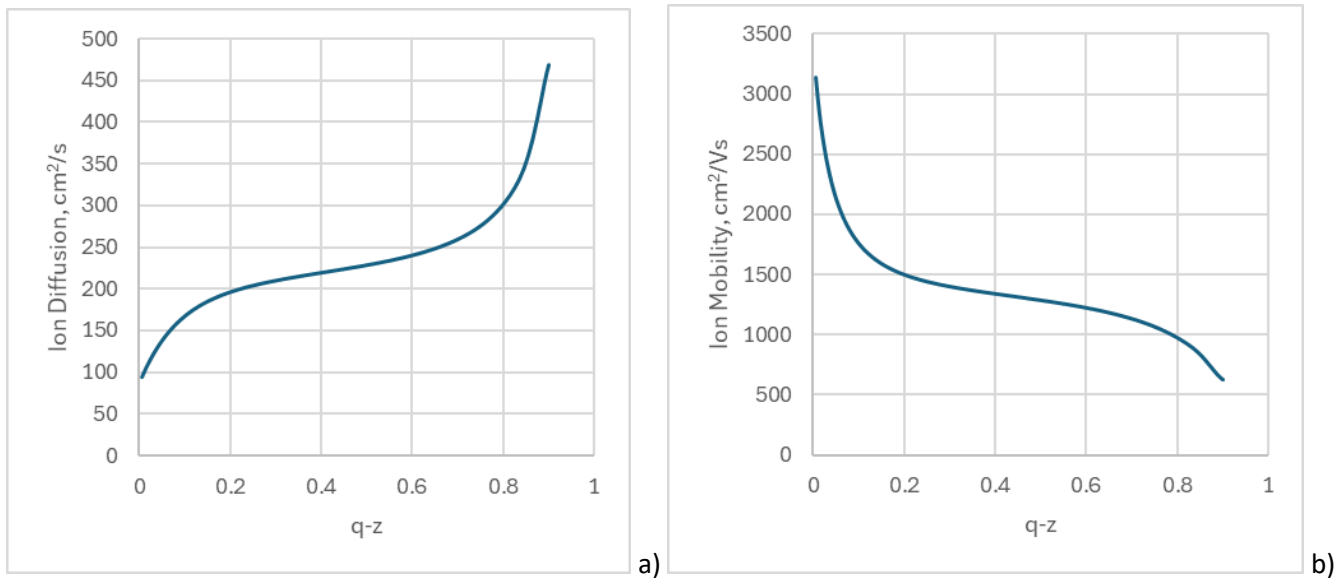

Figure 3. Ion a) diffusion and b) mobility in the HPIT vs. q-parameter. Modeling for M=16 Da, P=1 Torr, T=25C, m=100 Da, Rf=12 MHz, SR = 100 Da/ms, $z_0$=0.0635 cm, $r_0$ = 0.05 cm.

Generalized Einstein's relation connects mobility and diffusion $D = \varepsilon K$, where $\varepsilon$ is ion movement energy. Notice that ion mobility and diffusion depend on the ion energy and, therefore, on the electric field strength.

The mathematical model describing HPIT operation is based on the concept of ion cloud evolution in the PPW. The main approach is to find the characteristic ion cloud size as a function of the q-parameter, $s(q)$, dependent on the instrumental HPIT parameters, then calculate analytical HPIT parameters, which can be compared with experimental data.

Energy conservation law yields $s(q)$: $\varepsilon(q, s) + eU_R(s) = eU_C(q, s)$, where $\varepsilon$ is the ion movement energy; $U_R$ is the repulsive Coulomb potential; $eU_C(q, s) = c_1 F s^2$ is the pseudo-potential of the Ion Trap (see **Ion Trap theory** section for details); $c_1$ is the constant coefficient.

Through-gas **ion movement energy**[12] is determined by an electric field strength but not by a potential as in a vacuum. The mean energy of an ion propagated through the gas by the electric field is $\varepsilon = \varepsilon_T + \frac{1}{2} M v_d^2$, where $\varepsilon_T = 1.5kT$ is thermal energy; $v_d = KE$ is ion drift velocity, $E$ is electric field strength; $K = K_0/N$ is the ion mobility coefficient ($K_0$ is ion mobility under normal conditions, and $N$ is gas density).

Under HPIT conditions, strong repulsive ion-neutral interaction prevails, so the hard-sphere model of the ion mobility is valid. In the framework of this model, ion mobility depends on the ion energy as $K_0 = \frac{a}{\sigma_{HS}\sqrt{\varepsilon\mu}}$, where $\sigma_{HS} \sim (m^{1/3} + M^{1/3})^2$ is hard sphere diffusion cross-section; $a$ is constant coefficients. Substituting $K$ and solving the ion movement energy equation regarding $\varepsilon$ yields $\varepsilon = \varepsilon_T + \frac{b\langle E \rangle}{N\sigma_{HS}}\sqrt{\frac{M}{\mu}}$, where $b$ is the constant coefficient; $\langle E \rangle$ is the average electric field along the ion trajectory, which can be estimated by time and space averaging under the assumption of total trajectory entanglement. Given that time dependence is harmonic, ion cloud space distribution is quasi-Gaussian, and electric field amplitude ($E_{max}$) is proportional to the applied voltage V (and therefore to the q-parameter): $\langle E \rangle = \langle f(t) \rangle \frac{\langle En \rangle}{\langle n \rangle} \sim qz$. So, the ion movement energy is $\varepsilon = \varepsilon_T + c_2 q s$, where $c_2$ is the constant coefficient, $s$ is characteristic ion cloud size.

Trapped unipolar ions produce a space charge field directing outwards the trap center. It acts against the trapping field, directing towards the trap center, decreasing trapping efficiency. Under the assumption of the Gaussian distribution, the charge density is $\rho(r) = \frac{en}{s^3 \sqrt[3]{2\pi}} exp\left(-\frac{r^2}{2s^2}\right)$, where $n$ is the number of the trapped ions, $s$ is the characteristic ion cloud size. The solution of the Poisson equation $\nabla^2 U_R = -4\pi\rho$ yields repulsive **Coulomb potential** produced by the trapped ions $U_R(r) = \frac{n}{4\pi\varepsilon_0 r} erf\left(\frac{r}{\sqrt{2}s}\right)$. Finally, $U_R(s) = \frac{c_3}{s}$, where $c_3$ is the constant coefficient.

**Characteristic ion cloud size**, $s(q)$, is given by a solution of the equation $\varepsilon(q, s) + eU_R(s) = eU_C(q, s)$, which can be reduced to the cubic equation $c_1 F s^3 - c_2 q s^2 - \varepsilon_T s - c_3 = 0$. Formally, $s(0)$ and $s(0.908)$ rise to infinity, but trap walls limit the ion cloud spreading, so the actual size of the ion cloud can be estimated as $(s^{-2} + l^{-2})^{-0.5}$, where $l$ is the characteristic diffusion length. E.g. g for the Cylindrical Ion Trap $l = \left[\left(\frac{2.405}{r_0}\right)^2 + \left(\frac{\pi}{2z_0}\right)^2\right]^{-0.5}$, where $r_0$ is radius and $z_0$ is half-height.

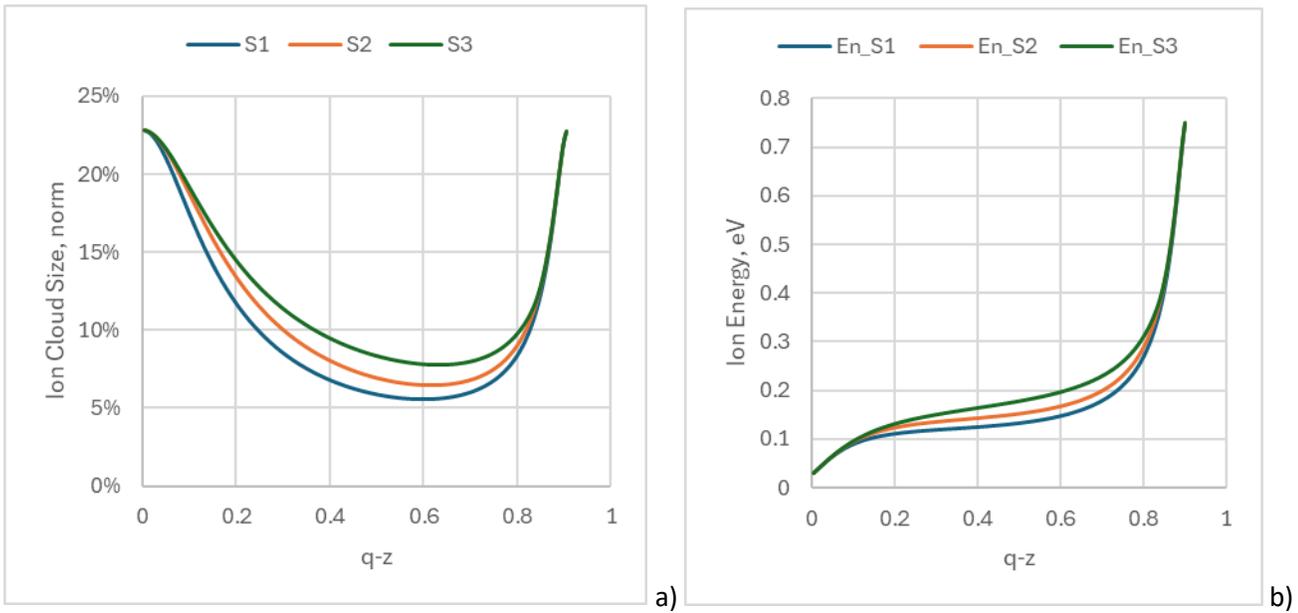

Figure 4. Ion a) cloud size and b) energy in the HPIT vs. q-parameter: S1 and En_S1 – simplified model; S2 and En_S2 – the thermal energy is taken into account; S3 and En_S3 – the Coulomb repulsion (3000 ions are trapped) is taken into account. Modeling for M=16 Da, P=1 Torr, T=25C, m=100 Da, Rf=12 MHz, SR = 100 Da/ms, $z_0$=0.0635 cm, $r_0$ = 0.05 cm.

HPIT is characterized by relatively strong ion scattering and high ion losses. The solution of the diffusion differential equation with zero boundary conditions (in τ-approximation) yields an estimation of the **ion losses** in the HPIT. The ion cloud in the conductive cavity without an electric field (q=0) diffuses towards the conductive walls to discharge there, and the ion number decreases exponentially with time $n(t) = n(0)\,exp(-t/\tau_0)$, where $\tau_0 = \frac{l^2}{D}$ is unconfined decay time.

The presence of the confinement potential ($U_c$) significantly reduces ion losses. The decay time increases exponentially[13] as $\tau = \tau_0\,exp(\gamma \tau_0/2)$, where focusing coefficient $\gamma = K\nabla^2 U_c \sim FK$. Substituting and reducing yields $\tau = \tau_0\,exp\left(c_1 l^2 \frac{F}{\varepsilon}\right)$, where $c_1$ is the constant coefficient; $\varepsilon$ is ion energy.

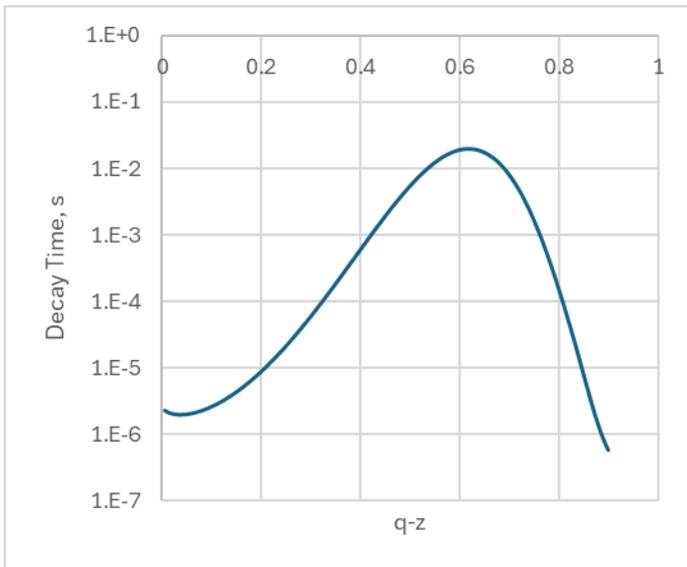

Figure 5 Ion density decay time in the HPIT vs. q-parameter. Modeling for M=16 Da, P=1 Torr, T=25C, m=100 Da, Rf=12 MHz, SR = 100 Da/ms, $z_0$=0.0635 cm, $r_0$ = 0.05 cm.

Thus, the HPIT ion trapping efficiency depends on the Stability Diagram operation point, the PPW depth, ion scattering (gas pressure), and trapping time. SD boundaries for the HPIT are the same as those for the low-pressure regular IT of the same geometry. The difference in the trapping efficiency can be explained by the fact that near the SD boundaries, PPW depth decreases, HPIT ion losses exponentially increase, and the range of effective trapping narrows. However, at very short trapping times, HPIT trapping efficiency tends toward that of the regular low-pressure IT. The shorter the trapping time and the lower the ion scattering (gas pressure), the closer the HPIT ion trapping efficiency across the SD to the regular LPIT.

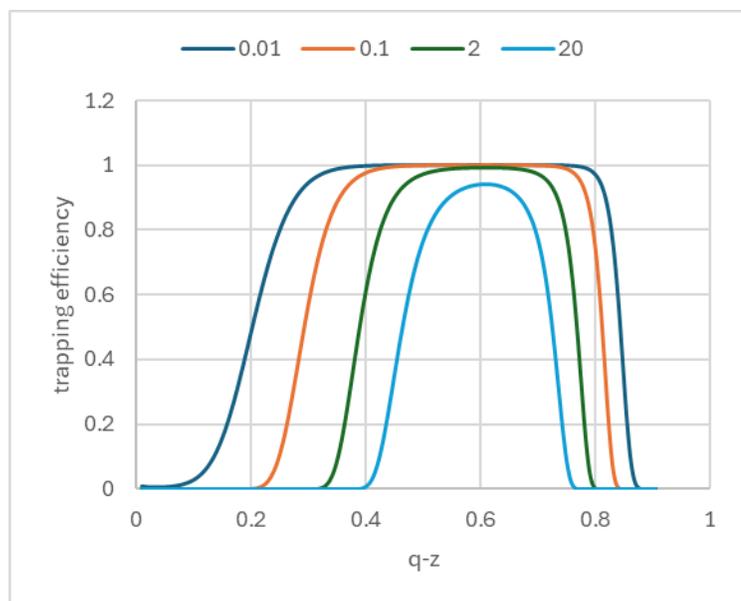

Figure 6. Trapping efficiency across the whole q range for the HPIT at different trapping times (0.01…10ms). Modeling for M=16 Da, P=1 Torr, T=25C, m=100 Da, Rf=12 MHz, SR = 100 Da/ms, $z_0$=0.0635 cm, $r_0$ = 0.05 cm.

## Results

Let's predict analytical HPIT characteristics (MS peak width, height, and position) in dependence on the instrumental parameters (IT size and geometry, gas pressure and composition, applied voltages and frequency, trapping time and scan rate) and ion properties (mass, diffusion, and mobility) to provide necessary data for the experimental verification of the model.

Existing cylindrical HPIT of the dimensions $r_0$ = 500 µm and $z_0$ = 650 µm operating at a pressure P=0.7…4 Torr of a neutral gas of mass M=4…40 Da analyzing ions of mass m=16…500 Da will be modeled to predict its analytical characteristics:

- MS peak height (PH),
- MS peak position (PP),
- MS peak width (PW),
- MS peak resolution (PW/PP).

in dependence on the following variables:

1. Trapping time and $q$ ($t_t$ and $q_t$);
2. Scan rate ($sr = dq/dt$);
3. Ion mass (m);
4. Neutral mass (M);
5. Number of the trapped ions (n);

6. Gas pressure (P);
7. Gas temperature (T);
8. RF frequency ($Rf = 2\pi\omega$).

HPIT analytical cycle starts from the **ion injection** followed by the **ion trapping**. Injected ions are trapped in the HPIT for some time (trapping time, $t_t$) at a certain q (trapping q, $q_t$). There is a different pattern of the trapping time and q dependencies for pulsed (HPIT with glow discharge ionizers) and continuous (HPIT with corona discharge or electrospray ionizer) ion injection.

In the case of the pulsed filling, HPIT is filled up to maximum capacity (defined by the IT PPW depth) and then loses ions exponentially during the trapping time. The remaining ions are ejected and detected as a mass-spectral peak. The number of ejected ions is $n_{ej} \sim n_{inj} \exp\left(-\frac{t_t}{\tau}\right)$, where $t_t$ is trapping time, $n_{inj}$ and $n_{ej}$ are the number of injected and ejected ions, $t_t$ is trapping time, $\tau$ is the decay time.

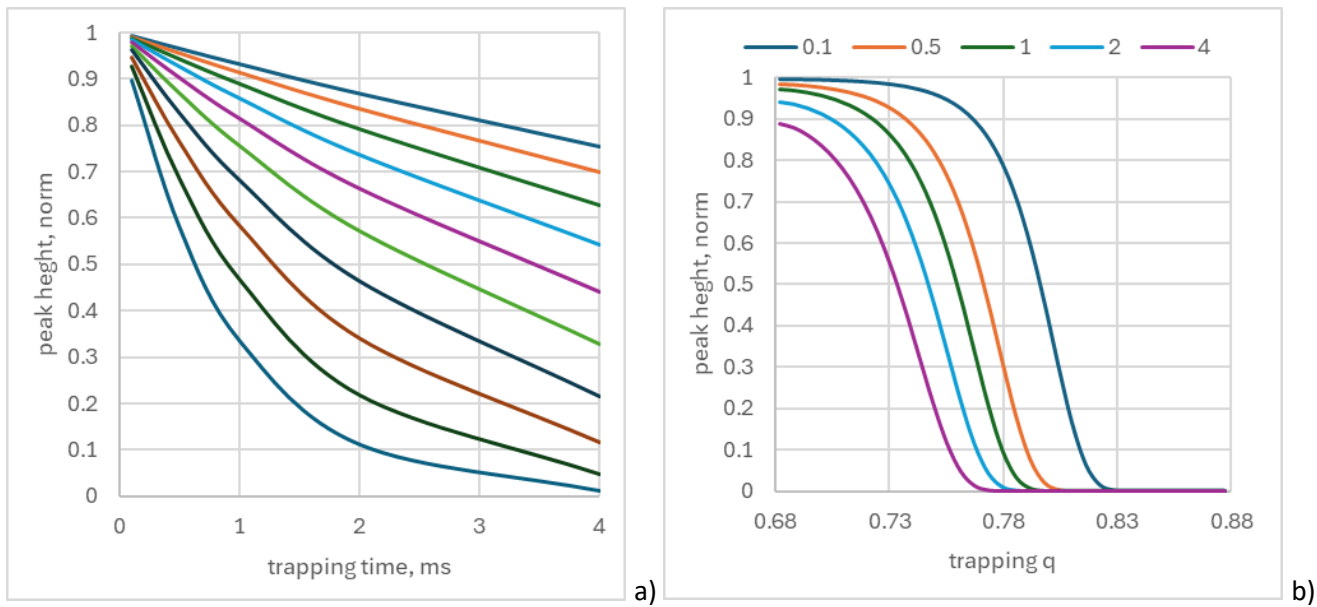

Figure 7. Normalized MS peak height vs. a) trapping time (trapping q = 0.75…0.8 as a parameter); b) trapping q (trapping time in is a parameter) for the **pulsed** injection. Modeling for M=16 Da, P=1 Torr, T=25C, m=100 Da, Rf=12 MHz, SR = 100 Da/ms, $z_0$=0.0635 cm, $r_0$ = 0.05 cm.

In the case of continuous filling, HPIT gains and loses ions simultaneously. Current-in ($J_{in}$) is constant, and current-out is $J_{out} \sim J_{in}\tau\left(1 - \exp\left(-\frac{t_t}{\tau}\right)\right)$, which results in a noticeable difference in the analytical characteristics.

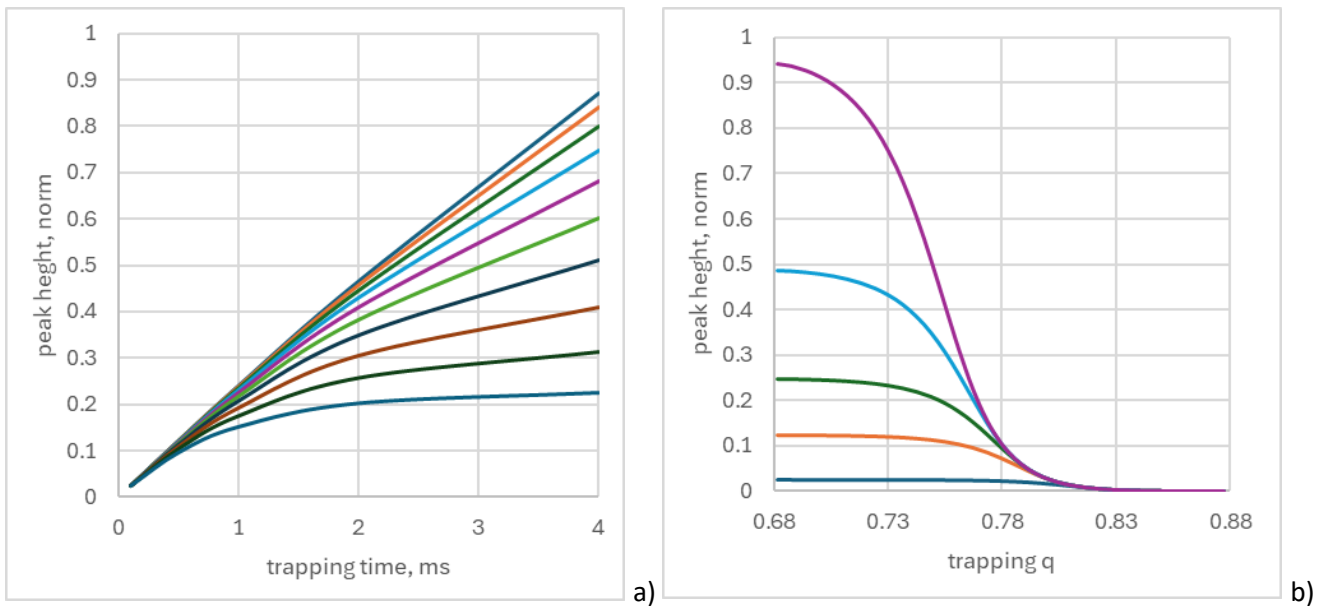

Figure 8. Normalized MS peak height vs. a) trapping time (trapping q = 0.75…0.8 as a parameter); b) trapping q (trapping time in is a parameter) for the **continuous** injection. Modeling for M=16 Da, P=1 Torr, T=25C, m=100 Da, Rf=12 MHz, SR = 100 Da/ms, $z_0$=0.0635 cm, $r_0$ = 0.05 cm.

The process of **ion ejection** can also be described in the framework of the proposed model. Let's suppose that a certain number ($n_0$) of ions is trapped at a certain $q$. Then $q$ increases linearly ($q(t)$ or RF scan) at a certain scan rate ($sr = dq/dt$). The decay coefficient $\tau(t)$ changes accordingly. The number of trapped ions is $n(t_s) = n_0 \, exp\left(-\int_0^{t_s} \frac{dt}{\tau(t)}\right)$, where $t_s$ is RF scan time. Ion current detected as a mass spectrum is $J = \frac{dn}{dt} = \frac{n(T)}{\tau}$. The model predicts a slightly asymmetric peak shape.

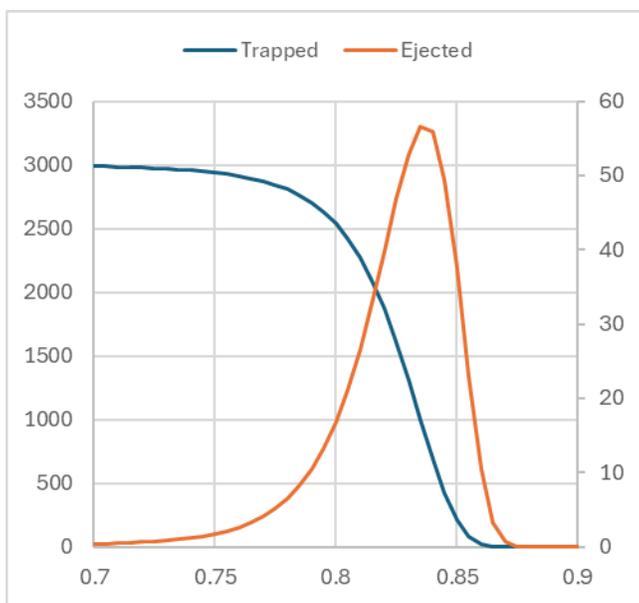

Figure 9. Number of the trapped and ejected ions vs. q-parameter. Modeling for M=16 Da, P=1 Torr, T=25C, m=100 Da, Rf=12 MHz, SR = 100 Da/ms, $z_0$=0.0635 cm, $r_0$ = 0.05 cm.

The next step is to model HPIT analytical characteristics. Modeling results are presented below in the form of the graphical dependencies of the HPIT analytical characteristics plotted against the HPIT parameters.

The parameters vary around the default values: neutral mass M=16 Da, pressure P=1 Torr, temperature = 25 C, ion mass m=100 Da, RF frequency =12 MHz, mass Scan Rate = 100 Da/ms, $z_0$=0.0635 cm, $r_0$ = 0.05 cm.

HPIT analytical characteristic corresponding to the default parameters are: peak position = 895 $V_{op}$, peak width at half maximum = 34.3 $V_{op}$, or 3.83 Da, a, mass resolution = 26.

Most important, non-obvious, and interesting results are placed in the main body of the article. All other results may be found in the **Appendix** section.

HPIT ion capacity is especially important for practical implementation. The proposed HPIT model shows that 600 trapped ions result in a 4% drop of the analytical characteristics. For using HPIT as a mass spectrometer core that allows replacing the electron multiplier (which requires a high vacuum for operation) by the Faraday's plate and electrometer for the detection of the ejected ions.

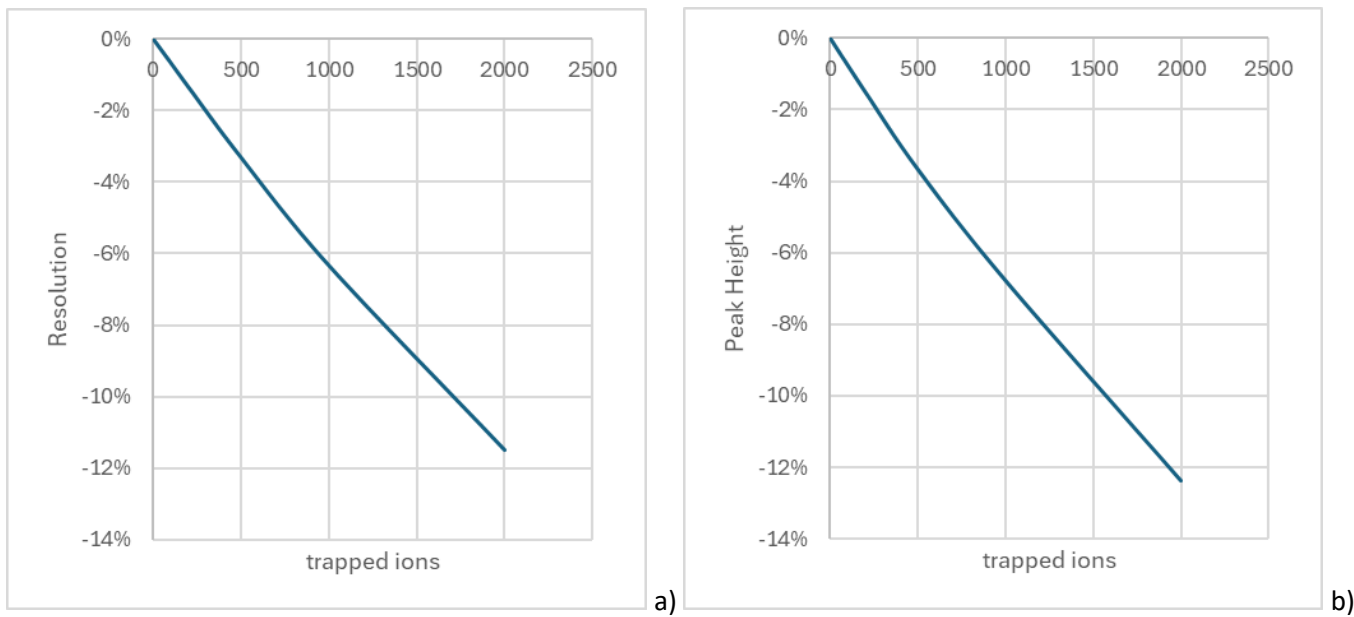

Figure 10. HPIT a) resolution and b) peak height vs. **number of trapped ions**.

The Mass Scan Rate appears to have a clear optimum (100 Da/ms) in terms of HPIT resolution (selectivity).

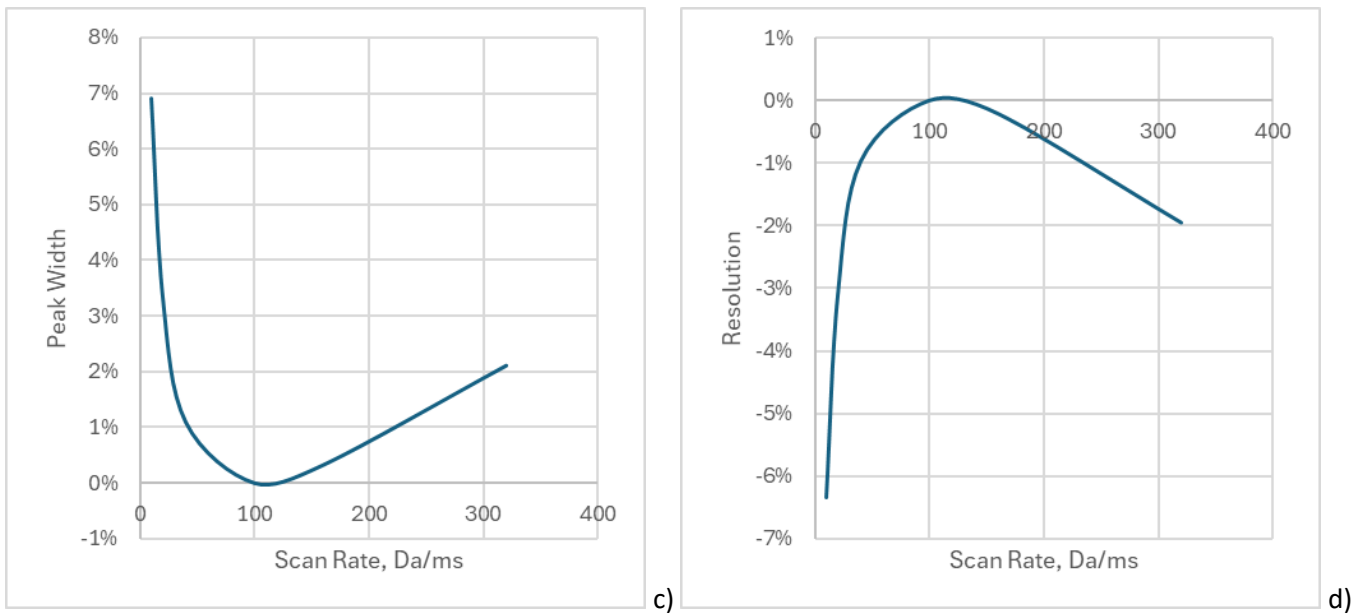

Figure 11. HPIT mass-spectrum peak a) width and b) resolution vs. **Scan Rate**.

Gas pressure 0.75 Torr corresponds to the maximal HPIT resolution (selectivity) and 1 Torr to the maximal peak height (sensitivity).

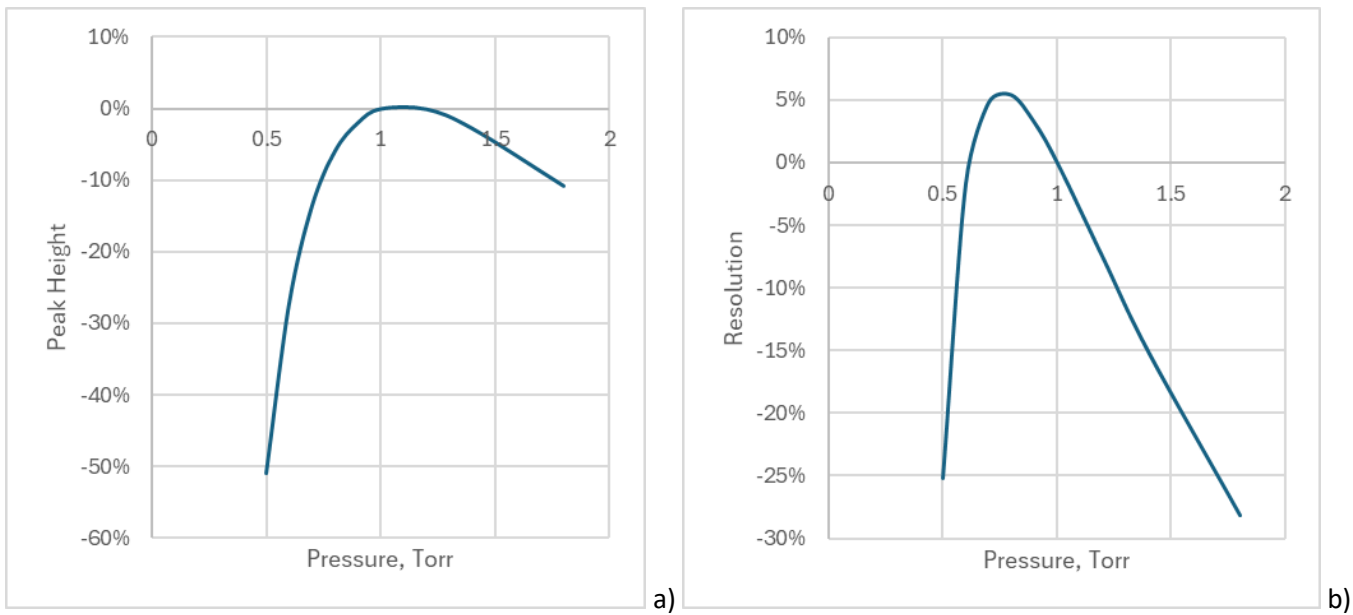

Figure 12. HPIT mass-spectrum peak a) height, b) resolution vs. **Pressure**.

HPIT with $z_0/r_0=1.13$ has the highest resolution (selectivity) and peak height (sensitivity).

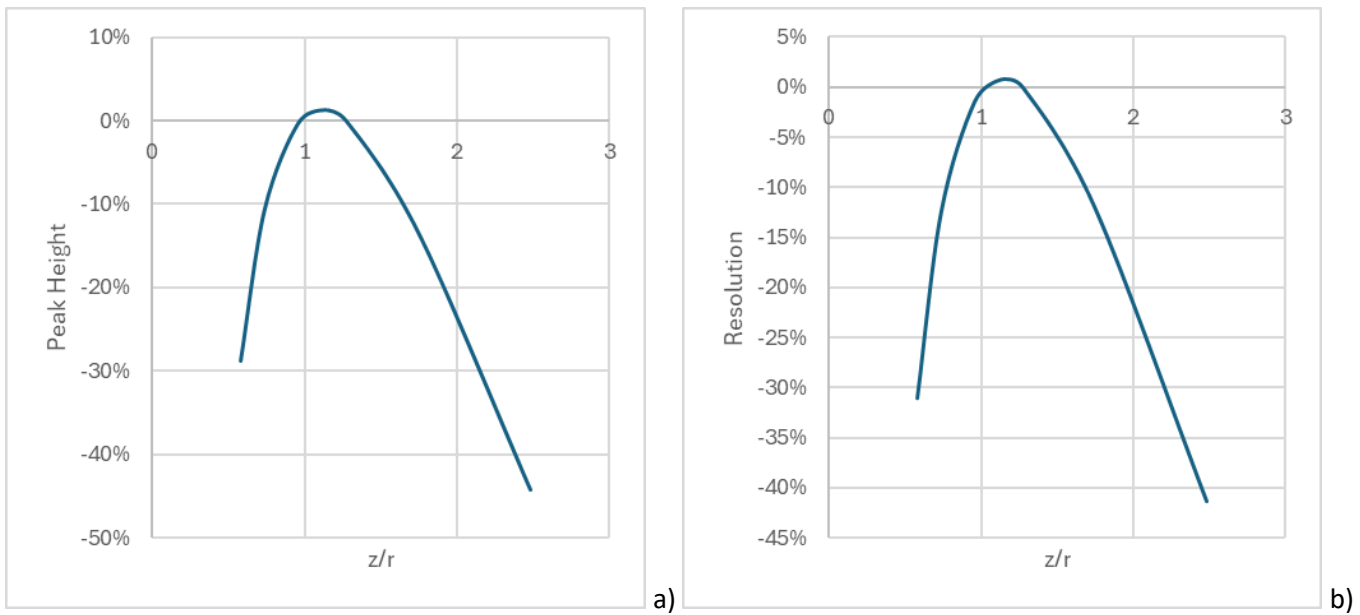

Figure 13. HPIT mass-spectrum peak a) height, b) resolution vs. HPIT **geometry ($z_0/r_0$)**.

All these theoretical predictions can be compared with the experimental data to validate the proposed model.

## Conclusion

A comprehensive model of the High-Pressure Ion Trap operation has been developed. This model explains and qualitatively describes the difference between the HPIT and regular Low-Pressure Ion Trap. In the framework of the model, the following problems have been solved:

- Through-gas ion movement in a potential well;
- Ion properties (energy, mobility, and diffusion) in the HPIT;
- HPIT ion capacity;
- Stability diagram for the HPIT;
- Ion ejection in the HPIT.

Based on the physical properties (ion and neutral mass; ion energy, mobility, and diffusion) and technical HPIT characteristics (dimensions, gas pressure, and applied voltages), analytical parameters (MS peak width, position, and height) of the HPIT were predicted.

This study should be helpful for the understanding and improvement of ion trapping RF devices at elevated pressure, aiming the development of a handheld High-Pressure Ion Trap Mass Spectrometer.

Finally, the differences between HPIT and regular IT are significant enough to believe that HPIT is a novel type of gas analytical device distinguished from regular mass spectrometers. In comparison with regular IT, HPIT is characterized by the following:

- Resolution is lower;
- Ion capacity is higher;
- Implementation is simpler;
- Size, weight, and power consumption are smaller.

# Appendix

Below are the rest of the modeling results in the form of the graphical dependencies of the HPIT characteristics variation plotted against the HPIT parameters variation. These dependencies are more or less obvious and/or trivial and are presented here for completeness. HPIT parameters vary around the default values: neutral mass M=16 Da, pressure P=1 Torr, temperature = 25 C, ion mass m=100 Da, RF frequency =12 MHz, mass Scan Rate = 100 Da/ms, $z_0$=0.0635 cm, $r_0$ = 0.05 cm. HPIT analytical characteristics vary around the default values: peak position = 895 $V_{op}$, peak width at half maximum = 34.3 $V_{op}$, or 3.83 Da, peak height = *0.014*$n_{tr}$ au ($n_{tr}$ is number of trapped ions), mass resolution = 26.

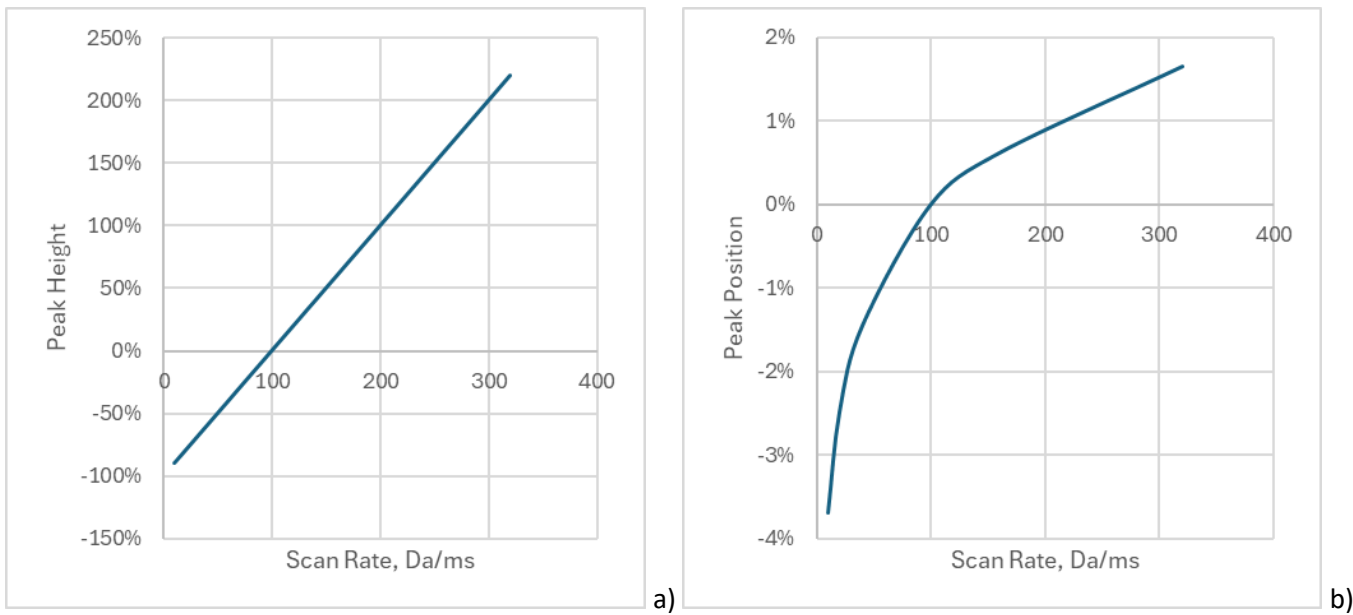

Figure A1. HPIT mass-spectrum peak a) height, b) position, vs. **Scan Rate**.

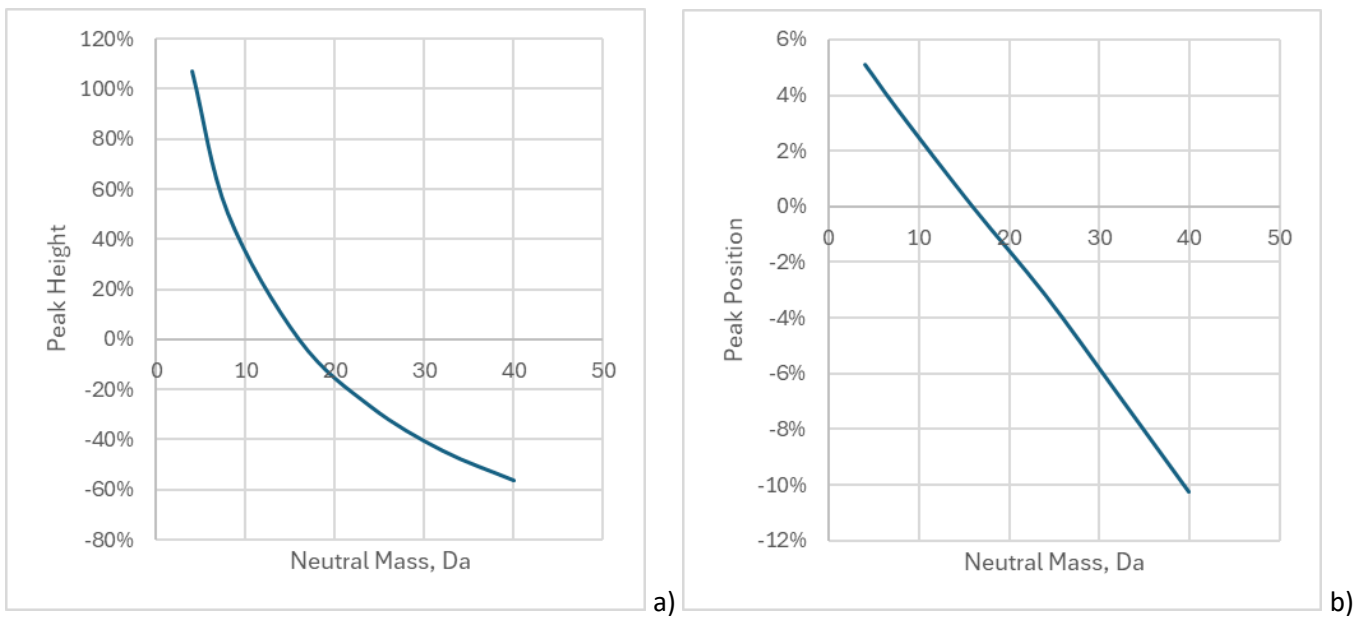

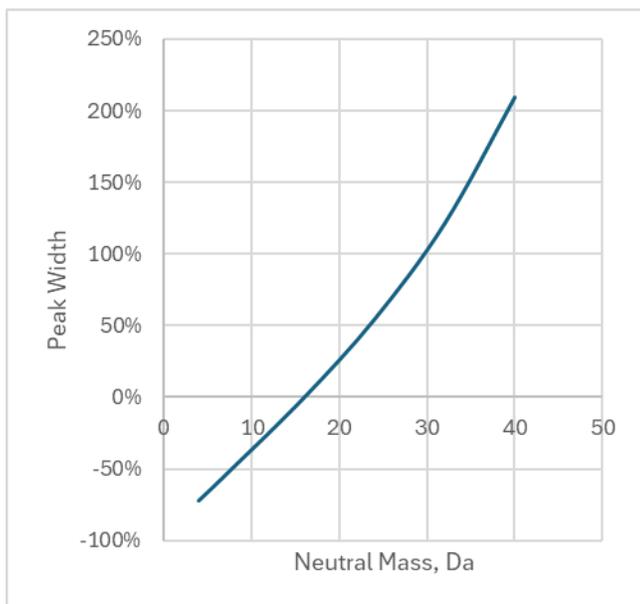
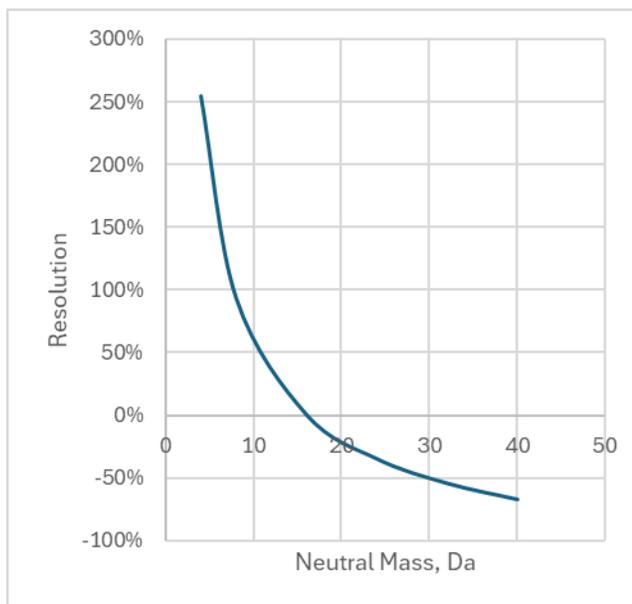

c)

d)

Figure A2. HPIT mass-spectrum peak a) height, b) position, c) width, and d) resolution vs. **Neutral Mass.**

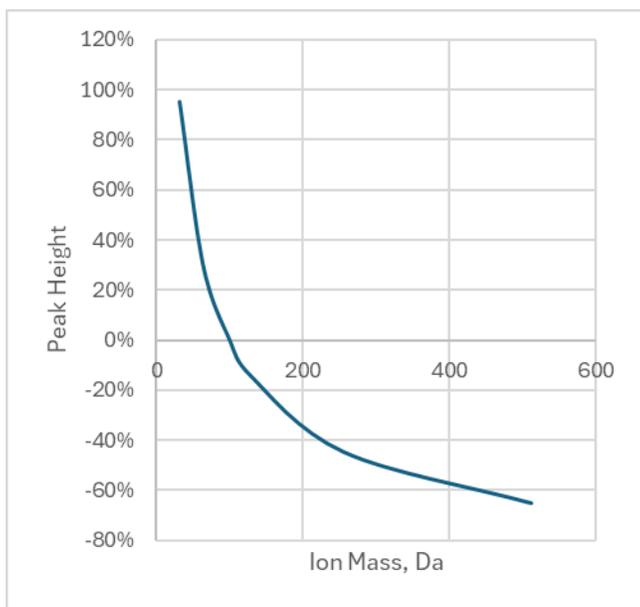
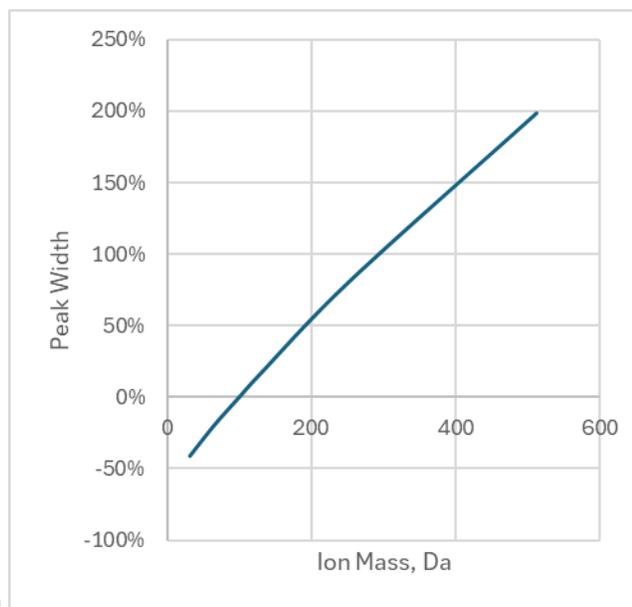

a)

b)

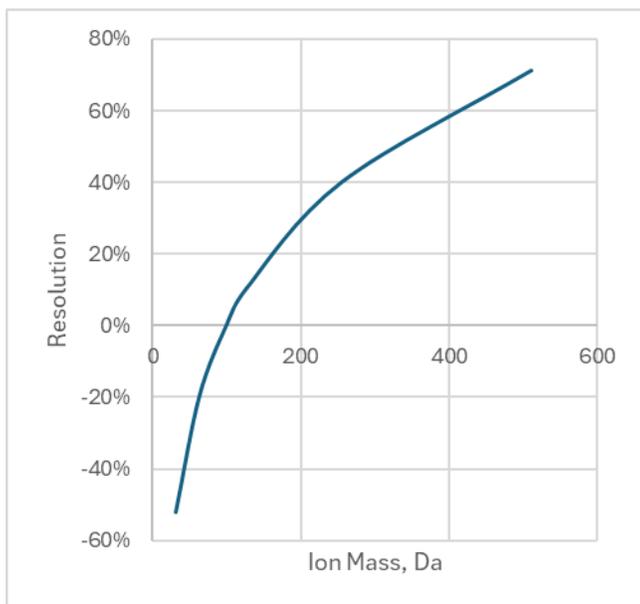
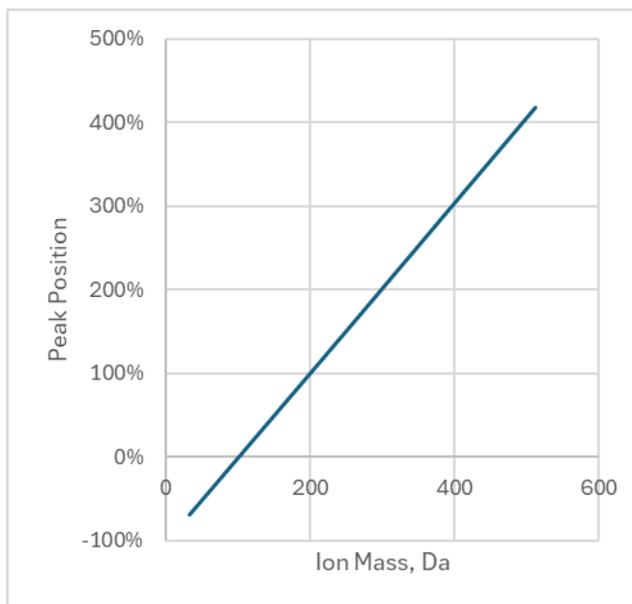

Figure A3. HPIT mass-spectrum peak a) height, b) width, c) resolution and d) position vs. **Ion Mass**.

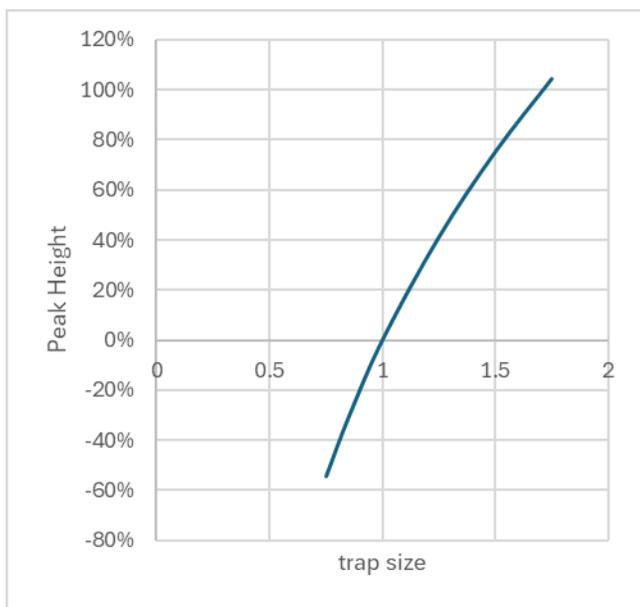
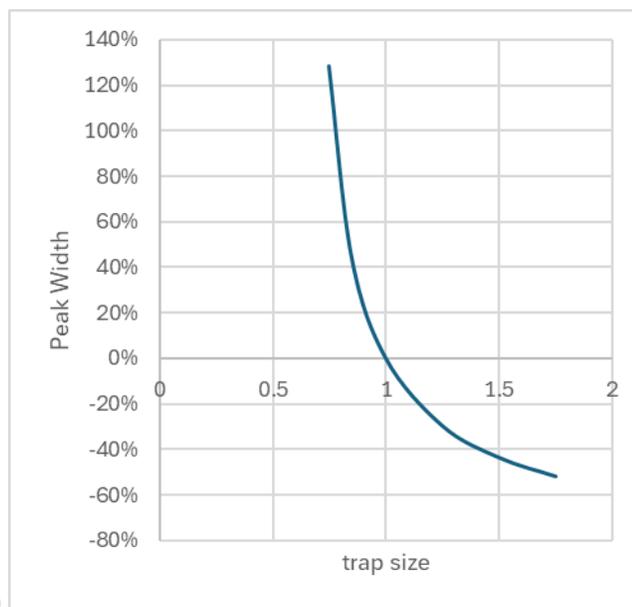

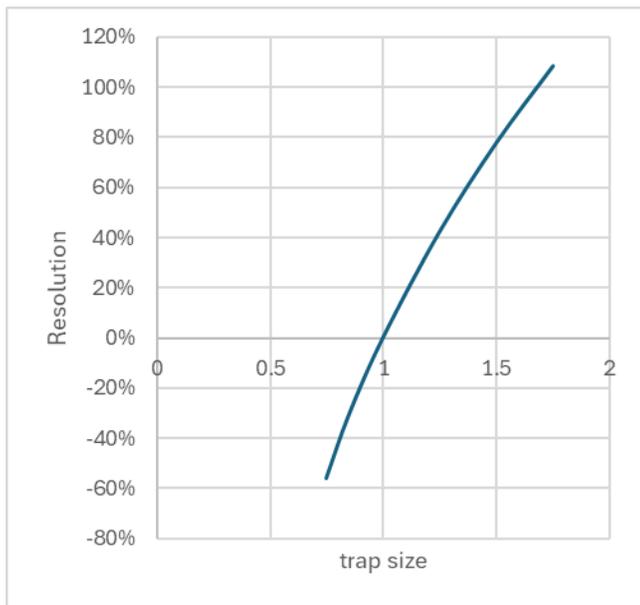
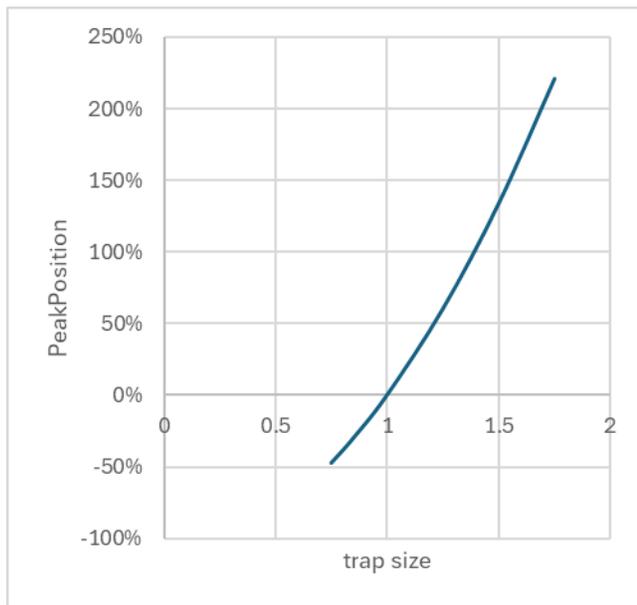
c)
d)

Figure A4. HPIT mass-spectrum peak a) height, b) width, c) resolution, and d) position vs. HPIT **size** (all HPIT dimensions change proportionally).

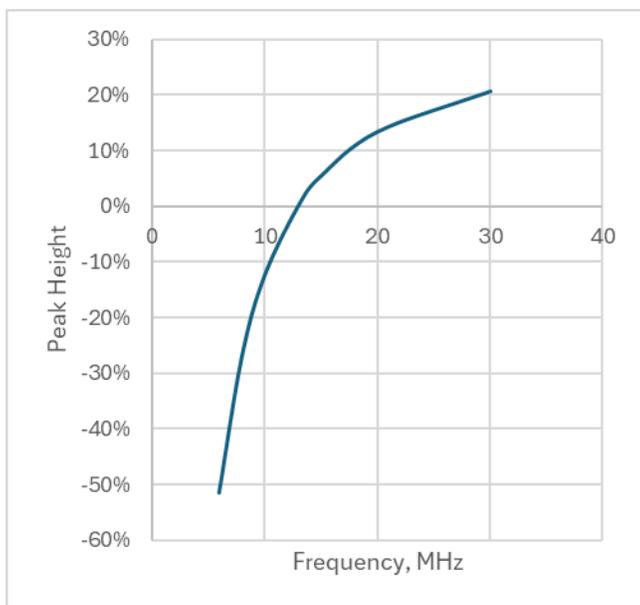
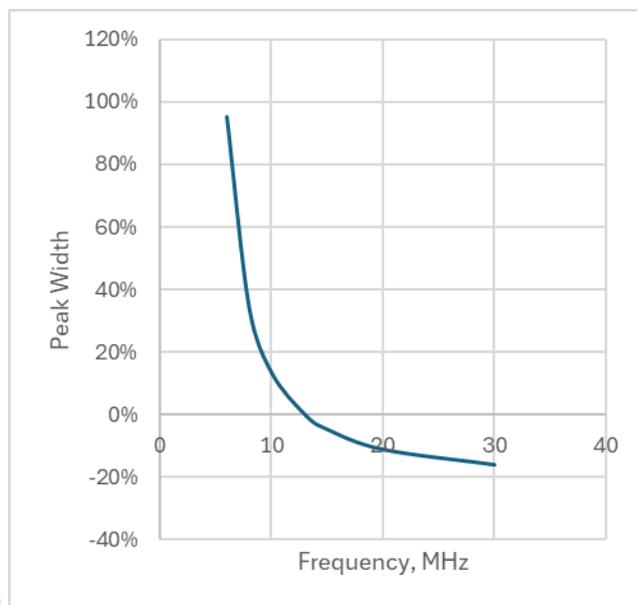
a)
b)

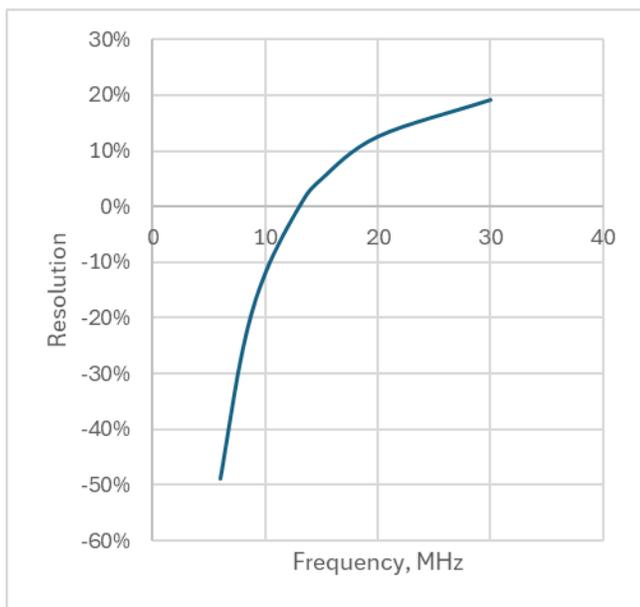
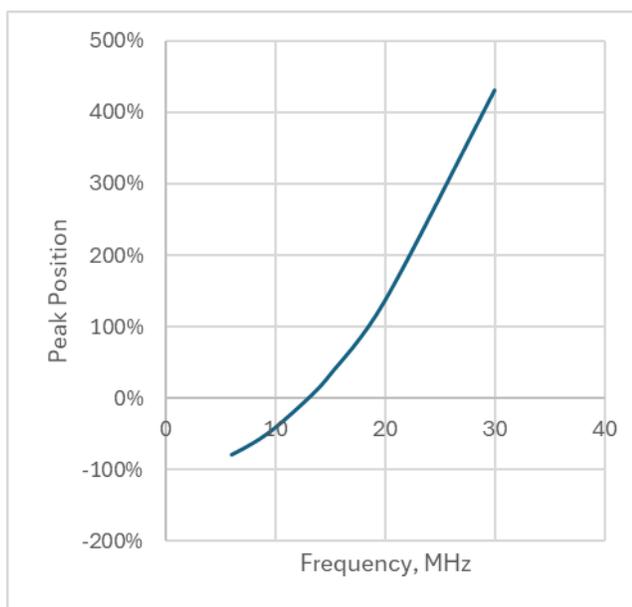

Figure A5. HPIT mass-spectrum peak a) height, b) width, c) resolution, and d) position vs. RF **Frequency**.

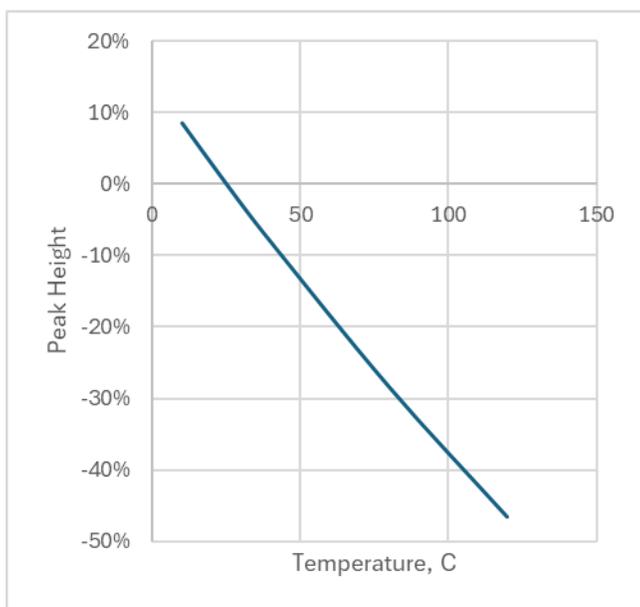
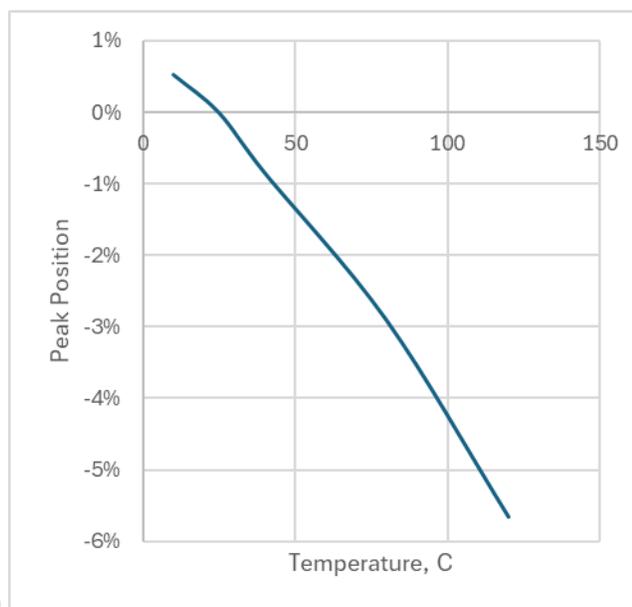

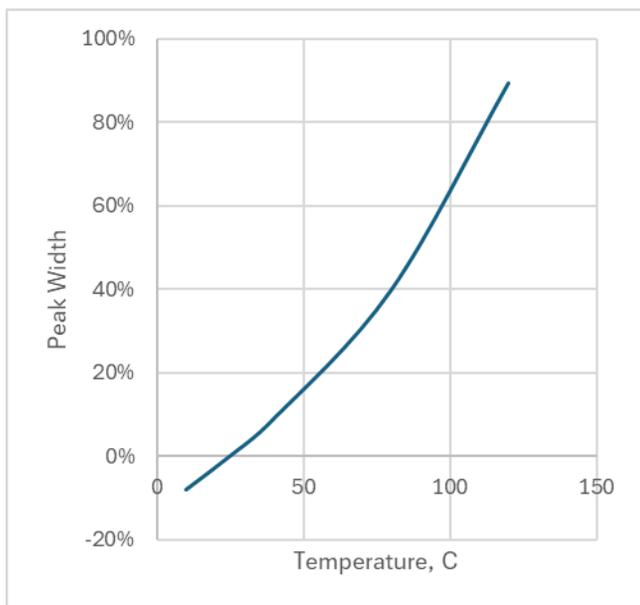
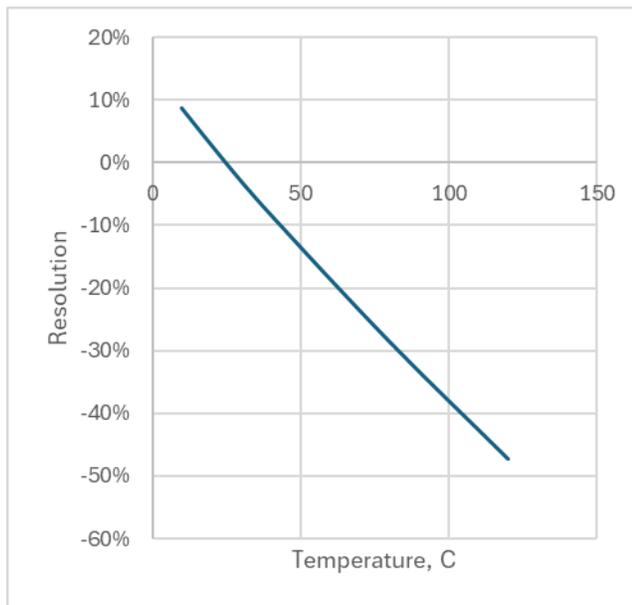

c)

d)

Figure A6. HPIT mass-spectrum peak a) height, b) position, c) width, and d) resolution vs. HPIT **temperature**.

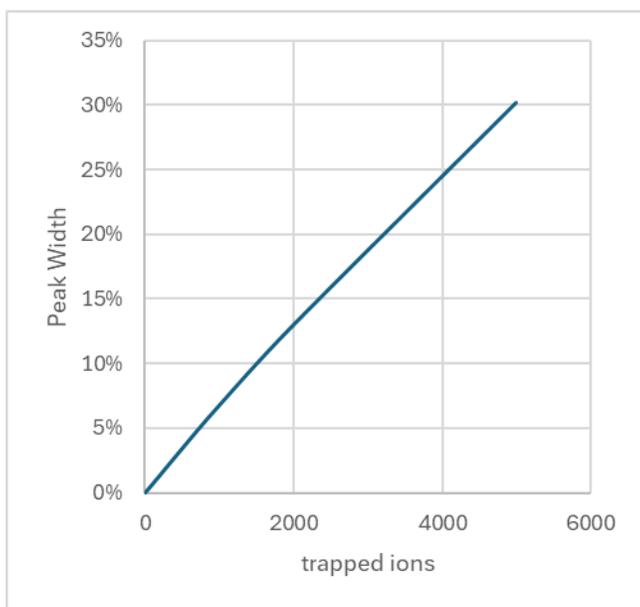
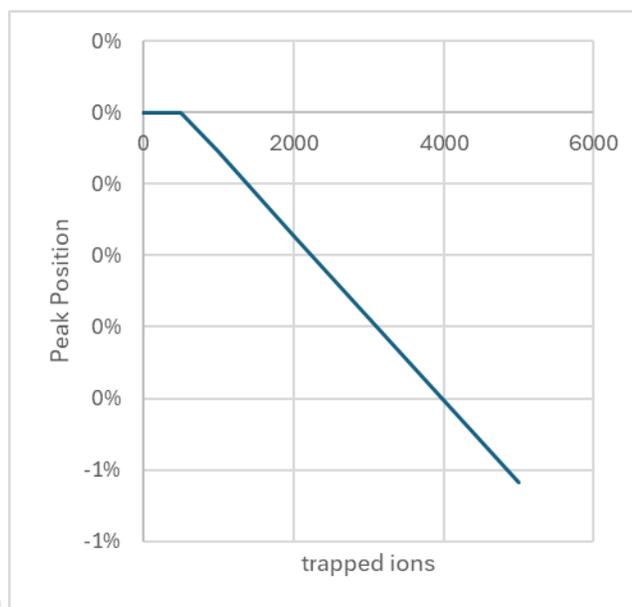

a)

d)

Figure A7. HPIT peak a) width and b) position vs. **number of trapped ions**.